\documentclass[cpp,a4paper,fleqn%
]{w-art}
\usepackage{times,cite,w-thm,amsmath}

\theoremstyle{plain}

\theoremstyle{definition}

\usepackage[]{graphicx}
\usepackage{epstopdf}
\begin{document}
\DOIsuffix{theDOIsuffix}
\Volume{xx}
\Month{xx}
\Year{xxxx}
\pagespan{1}{}
\Receiveddate{XXXX}
\Reviseddate{XXXX}
\Accepteddate{XXXX}
\Dateposted{XXXX}



\title{Simultaneous effect of an external magnetic field and gas-induced friction on the caging of particles in two-dimensional Yukawa systems}


\author[K. N. Dzhumagulova]{Karlygash N. Dzhumagulova\inst{1}\footnote{Corresponding author \quad E-mail:~\textsf{dzhumagulova.karlygash@gmail.com}}}
\address[\inst{1}]{IETP, Al-Farabi Kazakh National University, 71 Al-Farabi av., Almaty 050040, Kazakhstan}
\author [R. U. Masheyeva] {Ranna U. Masheyeva\inst{1}}
\author [T. S. Ramazanov] {Tlekkabul S. Ramazanov\inst{1}}

\author [G. Xia] {Guoxing Xia\inst{2}}
\address[\inst{2}]{School of Physics and Astronomy, University of Manchester, Manchester, United Kingdom}

\author [M.N. Kalimoldayev] {Maksat N. Kalimoldayev\inst{3}}
\address[\inst{3}]{Informational and Computational Technologies KS MES RK, Pushkin st., 125, Almaty, 050010, Kazakhstan}

\author[Z. Donk\'o]{Zolt\'an Donk\'o\inst{4}}
\address[\inst{4}]{Institute for Solid State Physics and Optics, Wigner Research Centre for Physics, Hungarian Academy of Sciences, Budapest, Hungary}


\begin{abstract}
We investigate the simultaneous effect of a static homogeneous external magnetic field  and a  background  gas  medium  on  the  quasi-localization  of  the  dust particles -- characterized  quantitatively  by  cage  correlation  functions -- in  strongly coupled two-dimensional  Yukawa  systems. We apply the Langevin  dynamics  computer simulation method in which the frictional and Lorentz forces are taken into account. Both the presence of the magnetic field or the friction originating from the background gas, when acting alone, increase the caging time. When present simultaneously, however, we find that their effects combine in a nontrivial manner and act against each other within a window of the parameter values.
\end{abstract}

\maketitle                   

\section{Introduction}
Strongly coupled plasmas are characterized by a pair-interaction potential energy that dominates over the average kinetic energy of the particles \cite{sccs}. Systems with this property appear in a wide variety of physical systems, as well in various laboratory settings and can be described by the ``one-component plasma'' (OCP) model, which considers explicitly only a single type of charged species and assumes an inter-particle potential that accounts for the presence and effects of the other type(s) of species. The non-polarizable form of the interaction potential is the Coulomb type, while the polarizable form is the Yukawa type; the corresponding systems are, respectively, quoted as Coulomb-OCP and Yukawa-OCP (YOCP). This latter type represents an important model system for dusty plasmas, e.g. \cite{Marlene,Bonitz,dusty1,dusty2}. In the simulations of these systems the background plasma environment is accounted for by the screening of the Coulomb potential. Moreover the presence of the gaseous environment may be taken into account by using the {\it Langevin simulation} approach \cite{LS1,LSV,LS2,LS3}, where two additional terms are incorporated into the equation of motion. One of the terms represents a friction by a homogeneous background, while the other term adds momentum to the particles in form of random kicks that change the direction of the motion. The action of these two terms can be balanced to reach a desired system temperature. In several studies, a simplified approach, which neglects the presence of the gaseous background, has been used to study the properties of idealized (frictionless) Yukawa liquids. This latter approach is justified for describing experiments conducted at very low pressures. The presence / absence of friction can significantly influence some of the properties of the systems, e.g. the lifetime of collective excitations.

The possibility of the presence of an {\it external magnetic field} has opened a new area of dusty plasma research. Interesting new effects (like the appearance of ordered structures) has been observed in the experiments \cite{S,TR,Ed}, which however, are closely related to the influence of the magnetic field on the discharge plasma itself that embeds the dust suspension. The reason for this is that due to the specific charge to mass ratios the electrons and ions become very strongly magnetized before any effect on the dust particles sets on. Understanding of the effects observed experimentally is difficult because of the very demanding computational needs -- simulation of dusty plasma experiments with strong magnetic fields is certainly an emerging area \cite{Peter}. To overcome the problem of the magnetization of the discharge plasma, another approach, based on the equivalence of the (magnetic) Lorentz force and the Coriolis force experienced in a rotating frame of reference was advised in \cite{HK1,HK2}. Implementation of this method and a successful identification of magnetoplasmons in a "quasi-magnetized" rotating dusty plasma was reported in \cite{MM}. 

Most of the computational studies have, meanwhile, concentrated on idealized systems, where the dust particles experience the effect of the magnetic field, whereas the surrounding medium is left unperturbed by this field. Many of the properties of such idealized systems have been studied by many-body (typically Molecular Dynamics) simulations. Collective excitations have been explored in \cite{coll,coll2}, the effect of the magnetic field on the diffusion was analyzed in \cite{diff1,diff2}, while heat transport was addressed in \cite{heat}.

Many of the properties of the strongly coupled complex plasmas are strongly related to one of its outstanding features, the {\it quasi-localization of the particles} in these systems \cite{GK2000}:
the particles oscillate in the local wells of the potential surface, which changes due to the diffusion of the particles on a timescale that can be significantly longer as compared to the timescale of oscillations. 
A mathematical framework based on tracking the surroundings of individual particles has been developed in \cite{Rabani}: the duration of the localization can be quantified by means of the so-called "cage correlation functions". 
The effect of a static uniform external magnetic field on the cage correlation functions in frictionless 2-dimensional Yukawa systems has been investigated in \cite{cagemagnetic}, while the effect of the friction force, induced by the presence of the buffer gas, has been addressed in \cite{cagefriction}. Here, our aim is to study the {\it simultaneous effect of the magnetic field and the friction}, on the cage correlation functions. The interplay of these two effects is an open question, that can only be answered by a systematic parametric study, due to the inherent non-linearity of the system under investigation. Our studies are based on Langevin dynamics simulation into which a proper description of the movement of the particles under the influence of an external magnetic field is incorporated \cite{Langevin_sim,diff3,diff4}.

Our numerical integration scheme of the particles' equations of motion follows the approach of \cite{spreiter}, which takes into account the external magnetic field in the expansion of positions and velocities in the Taylor series. In \cite{computation}, we introduced the friction force into the Velocity Verlet scheme, which is used in the present simulations. The scheme has been verified via comparisons of the cage correlation functions obtained in the limiting cases, when the friction force or the Lorentz force tends to zero. 

The model and the computational methods are described in section 2, while the results are presented in section 3. A brief summary is given in section 4.

\section{Model and computational method}

We adopt the following form for the potential that results from the mutual interaction of the particles and the screening property of the surrounding plasma environment:
\begin{equation}
\label{eq:potential}
\phi(r) = \frac{Q}{4 \pi \varepsilon_0} \frac{\exp(-r/\lambda_{\rm D})}{r},
\end{equation}
where $Q$ is the charge of the particles and $\lambda_{\rm D}$ is the screening (Debye) length. 

We study a two-dimensional (2D) system, the particles move in the $(x,y)$ plane and the magnetic field is assumed to be homogeneous and directed perpendicularly to the layer of the particles, i.e., ${\bf B}=(0,0,B)$.
The equation of motion of the particles (given here for particle $i$) is
\begin{equation}
\label{eq:motion}
m \ddot{{\bf r}}_i(t) = \sum_{i \neq j}{\bf F}_{ij}(r_{ij})+{Q}[{\bf v}_i{\bf\times B}]- \nu ~m {\bf v}_i(t) + {\bf F}_{\rm Br},
\end{equation}
where the first term on the right hand side gives the sum of inter-particle interaction forces (to be computed for $(i,j)$ particle pairs that are separated by a distance $r_{ij}$), the second  is  the  Lorentz  force, the  third  term represents  the  friction  force (proportional to the particle velocity, $\nu$ is the friction coefficient of the dust particles in the  background gaseous  environment), while the fourth term represents an additional randomly fluctuating ``Brownian'' force that models the random kicks of the gas atoms on the dust particles. 

The ratio of the inter-particle potential energy to the thermal energy is expressed by the coupling parameter
\begin{equation}
\label{eq:gamma}
\Gamma = \frac{Q^2}{4 \pi \varepsilon_0 a k_{\rm B} T},
\end{equation}
where $T$ is temperature and $a = (1/\pi n)^{-1/2}$ is the  two-dimensional Wigner-Seitz radius with $n$ being the areal number density of the particles. We introduce the screening parameter $\kappa = a / \lambda_{\rm D}$. The strength of the magnetic field is expressed in terms of
\begin{equation}
\label{eq:beta}
\beta = \Omega / \omega_{\rm p},
\end{equation}
where $\omega_{\rm p}=\sqrt{n Q^2/2 \varepsilon_0 m a}$ is the nominal 2D plasma frequency and $\Omega$ is the cyclotron frequency.
The strength of the friction is defined by the dimensionless parameter
\begin{equation}
\label{eq:theta}
\theta = \nu / \omega_{\rm p}.
\end{equation}
So, the system is fully characterized by four parameters: $\Gamma$, $\kappa$, $\beta$ and $\theta$.

We apply the Langevin dynamics (LD) simulation method to describe the motion of the particles governed by the  equation of motion given above. To integrate this equation, a new numerical scheme based on the Taylor expansion of the particle acceleration and velocity, followed by the correct choice of all the terms that are not higher than $O((\Delta t)^2)$  is  used, in which the time step does not depend on the magnitude of the magnetic field. This scheme was obtained by applying the same technique developed in \cite{spreiter}, but takes into account the friction  force \cite{computation}. We obtained the following equations for the positions and velocities of the particles without taking into account $\bf F_{Br}$ (which can be added subsequently):

\begin{eqnarray}
\label{eq:coordx}
r_x(t+\Delta t)=r_x(t)-\frac{1}{(\Omega^2+\nu^2)}\big[(\nu v_x(t)+\Omega v_y(t))(\exp(-\nu \Delta t) \cos(\Omega \Delta t)-1) +\nonumber \\+(\nu v_y(t) -\Omega v_x(t)) \exp(-\nu \Delta t)\sin (\Omega \Delta t) \big]+ \nonumber \\+ \frac{1}{(\Omega^2+\nu^2)^2} \big[C(\Omega \Delta t)((\nu^2-\Omega^2)a^c_x(t)+2 \nu \Omega a^c_y(t))+\nonumber \\+S(\Omega \Delta t)((\nu^2-\Omega^2)a^c_y(t) -2 \nu \Omega a^c_x(t))\big]
\end{eqnarray}
$r_y(t+\Delta t)$ can be obtained from (\ref{eq:coordx}) by replacing $x \rightarrow y$ and $\Omega \rightarrow -\Omega$. Here $a^c$ is the part of the acceleration which does not depend on the velocities, furthermore, 
\begin{equation}
\label{eq:S}
S(\Omega \Delta t)\equiv \exp(-\nu \Delta t)\sin(\Omega \Delta t)-\Omega \Delta t
\end{equation}
and
\begin{equation}
\label{eq:C}
C(\Omega \Delta t) \equiv \exp(-\nu \Delta t)\cos(\Omega \Delta t)-1+\nu \Delta t.
\end{equation}
The velocity components are given as:
\begin{eqnarray}
\label{eq:velocityx}
v_x(t+\Delta t)=\exp(-\nu \Delta t)(v_x(t)\cos(\Omega \Delta t)+v_y(t)\sin(\Omega \Delta t)) +\nonumber \\+ \frac{1}{\Omega^2+\nu^2} \big[ \exp(-\nu \Delta t)(\Omega \sin(\Omega \Delta t)-\nu \cos(\Omega \Delta t))a^c_x(t)+\nu a^c_x(t) -\nonumber \\ -\exp(-\nu \Delta t)(\Omega \cos(\Omega \Delta t)+\nu \sin(\Omega \Delta t))a^c_y(t)+\Omega a^c_y(t) \big]+\nonumber \\ +\frac{1}{(\Omega^2+\nu^2)^2} \big[ \{ \exp(-\nu \Delta t)((\nu^2-\Omega^2)\cos(\Omega \Delta t)-2 \nu \Omega \sin(\Omega \Delta t))+\nonumber \\+(\Omega^2-\nu^2)+(\Omega^2+\nu^2)\nu \Delta t \} \frac{d}{dt} a^c_x(t) +\{ \exp(-\nu \Delta t)((\nu^2-\Omega^2)\sin(\Omega \Delta t)+\nonumber \\+2 \nu \Omega \cos(\Omega \Delta t))-2 \nu \Omega + (\Omega^2+\nu^2)\nu \Delta t \} \frac{d}{dt} a^c_y(t) \big]
\end{eqnarray}
$v_y(t+\Delta t)$ can be obtained from (\ref{eq:velocityx}) by replacing $x \rightarrow y$ and $\Omega \rightarrow -\Omega$

\vspace{0.5cm}

We investigate the localization of the particles characterized by the cage correlation function by using the method of\cite{Rabani}, that allows to track the changes surroundings of individual particles. We use a generalized neighbour list $\ell_i$ for particle $i$, $\ell_i = \{f_{i,1},f_{i,2},...,f_{i,N} \}$. ($f_{i,i}$ is excluded from the neighbor list, i.e. only "real" neighbors are taken into account.) Due to the underlying sixfold symmetry of the system considered here we always search for the six closest neighbours of the particles and the $f$-s corresponding to these particles are set to a value 1, while all other $f$-s are set to 0.

The similarity between the surroundings of the particles at $t=0$ and $t>0$ is measured by the {\it list correlation function} (defined as the normalized scalar product of the list correlation functions at times $t$ and $0$):
\begin{equation}
C_{\ell}(t) = {{\langle \ell_i(t) \cdot \ell_i(0) \rangle} \over
{\langle \ell_i(0)^2 \rangle}},
\label{eq:listcorr}
\end{equation}
where $\langle ~~ \rangle$ denotes averaging over particles and initial times. Obviously, $C_{\ell}(t=0) = 1$, and $C_{\ell}(t)$ is a monotonically decaying function.

The number of particles that have left the original cage of particle $i$ at time $t$ can be determined as
\begin{equation}
N_i^{\rm out} (t) = \ell_i(0)^2 - \ell_i(0) \cdot \ell_i(t).
\label{eq:n}
\end{equation}
Here, the first term gives the number of particles around particle $i$ at $t$ = 0 that equals to six in our case. The second term gives the number of ``original'' particles that remained in the surrounding after time $t$. The {\it cage correlation function} $C_{\rm cage}^{c}(t)$ is obtained by averaging over particles and initial times the function
$\Theta (c - N_i^{\rm out})$, i.e.
\begin{equation}
C_{\rm cage}^{c}(t) =\langle\Theta(c - N_i^{\rm out} (0,t)) \rangle,
\label{eq:cagecorr}
\end{equation}
where $\Theta$ is the Heaviside function. We compute the cage correlation functions for $c=3$, and take the definition of the {\it "caging time"} introduced in \cite{caging}, according to which $t_{\rm cage}$ is defined as the time when $C_{\rm cage}^{3}$ decays to a value 0.1.

The number of simulated particles is fixed at $N=1000$ that move within a quadratic simulation box. The positions of the particles are chosen randomly at the initialization of the simulations, their velocities are sampled from a Maxwellian distribution with a temperature that corresponds to the value of specified $\Gamma$. During the initial phase of the simulations the system is thermalized, but thermostation is stopped before the data collection phase starts.

\section{Results}

Below we present the results of our simulations obtained for the cage correlation functions under the conditions of  the simultaneous presence of the external magnetic field and the friction imposed by the background gaseous environment. In order to use the same numerical scheme throughout our work, we had to use finite values of the magnetic field and the friction coefficient. The field-free and/or frictionless cases are approximated by using very small values of these coefficients ($10^{-6}-10^{-5}$) in the simulations. The results obtained this way approximate the "true" $\beta=0$ and/or $\theta=0$ results well within the statistical noise of the results. Nonetheless, at the presentation of the results we give the precise (low) values of these coefficients used in the respective simulations.

\begin{figure}[h!]
\centering
\includegraphics[width=0.8\textwidth]{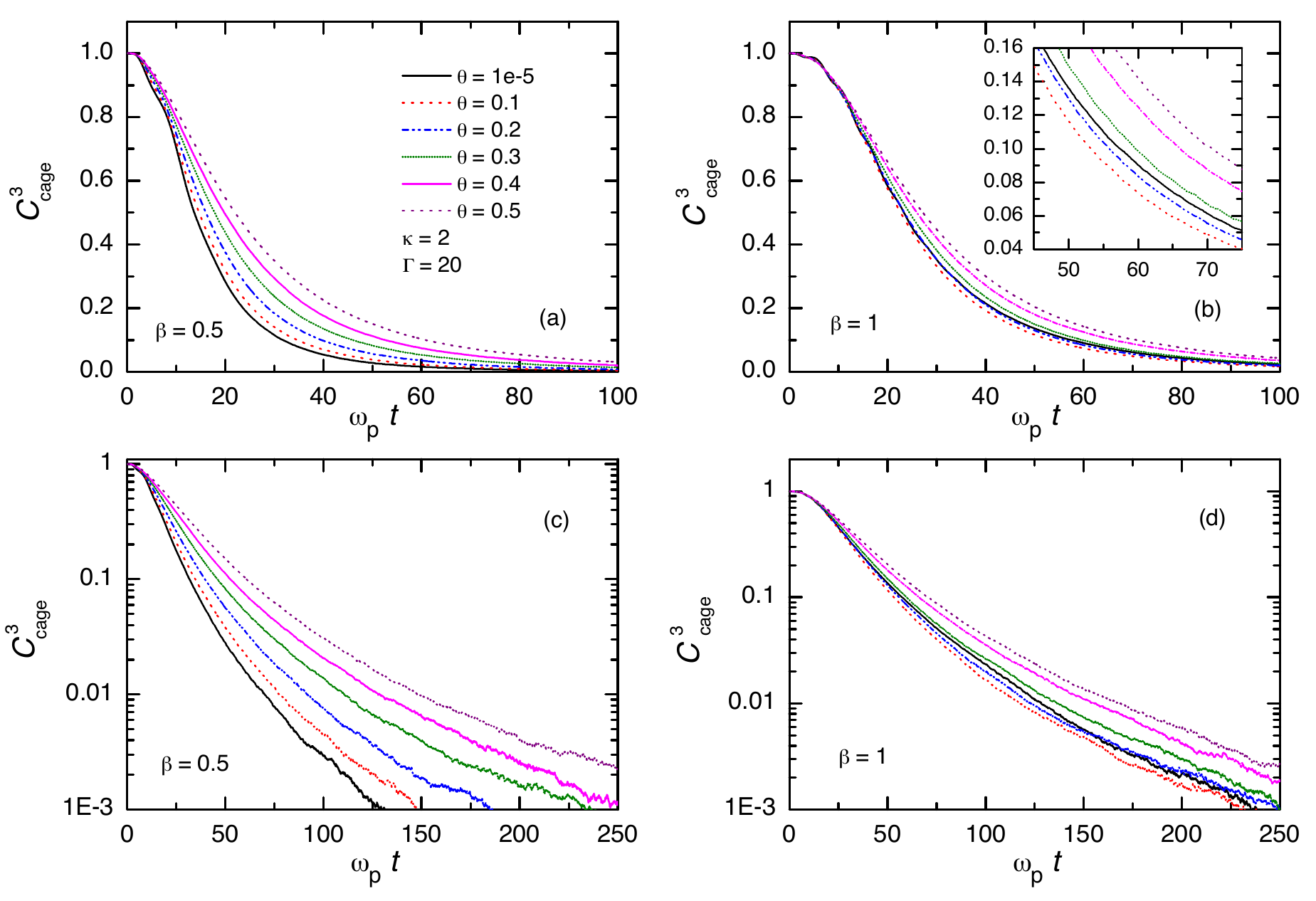}
\caption{\label{fig:cagecorr1} 
Cage correlation functions for $\Gamma=20$ and $\kappa=2$ for (a) $\beta$ = 0.5 and (b) $\beta=1$, for a wide range of the friction coefficient $\theta$. The legend shown in (a) also holds for panel (b). The inset in (b) zooms at the region when the correlation functions cross the $C_{\rm cage}=0.1$ line (at times that correspond to the caging time.) Panels (c) and (d) show the same functions with semi-logarithmic scales to allow a better observation of the long-time behavior.}
\end{figure}

\begin{figure}[h!]
\centering
\includegraphics[width=0.8\textwidth]{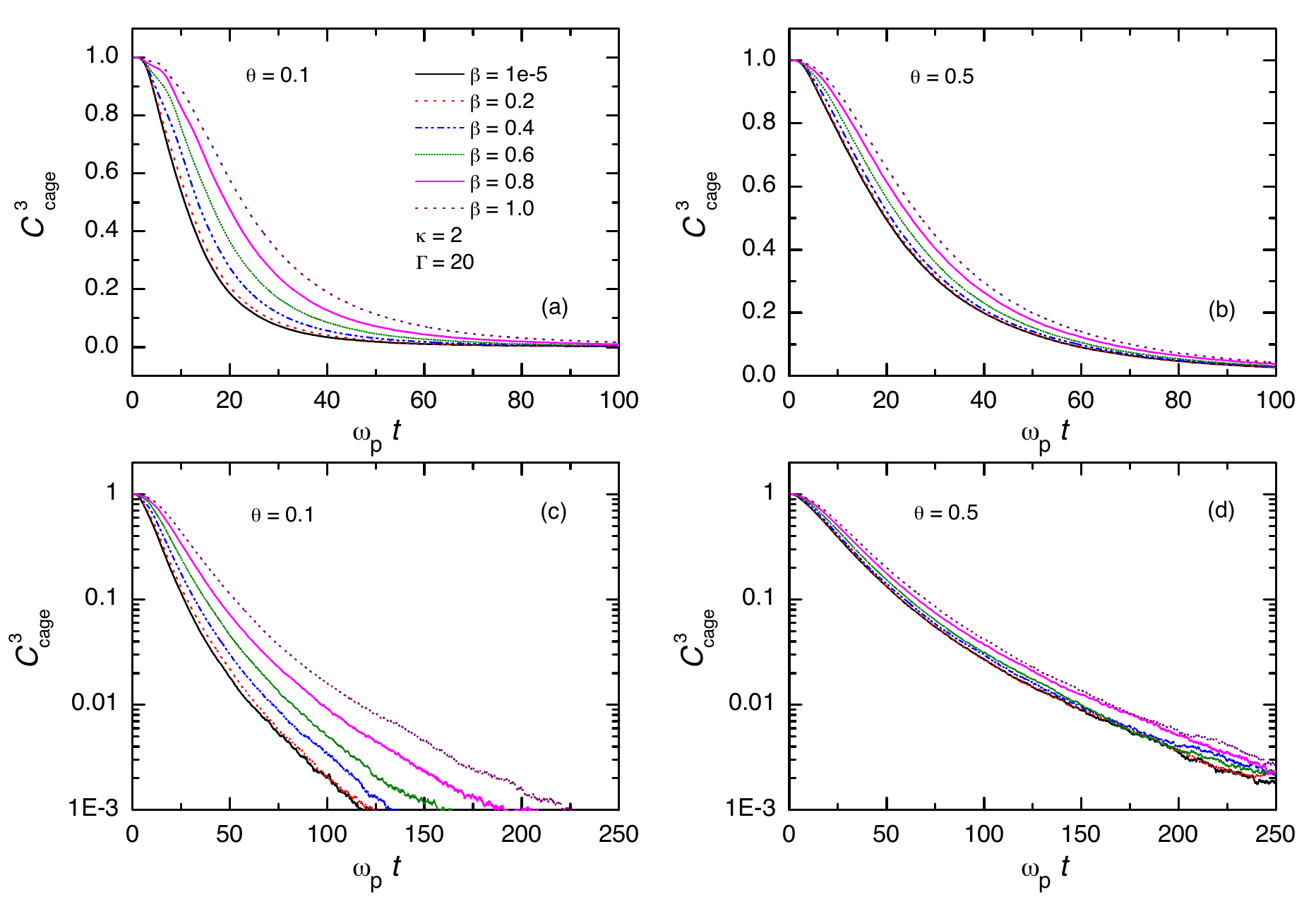}
\caption{\label{fig:cagecorr2} 
Cage correlation functions for $\Gamma=20$ and $\kappa=2$ for (a) $\theta$ = 0.5 and (b) $\theta=1$, for a wide range of the reduced magnetic field $\beta$. Panels (c) and (d) show the same functions with semi-logarithmic scales to allow a better observation of the long-time behavior.}
\end{figure}

Figure~\ref{fig:cagecorr1}(a) shows the $C_{\rm cage}^{3}(t)$ functions obtained at fixed system parameters $\kappa = 2$ and $ \Gamma = 20$ at a magnetic field $\beta$ = 0.5, with the friction coefficient $\theta$ scanned over the domain between $10^{-5}$ (representing a case with vanishing friction) and 0.5 (representing a case with high friction). In panel (b) of the same figure another set of data are presented for a stronger magnetic field of $\beta$ = 1. One plasma oscillation period, disregarding the effect of the magnetic field, corresponds to $\omega_{\rm E} T \sim 2 \pi$,  where $\omega_{\rm E}$ is the Einstein frequency \cite{review} that reflects the slowing down of the dynamics due to the effect of the screening. At the given value of the screening parameter, $\kappa=2$, we find $\omega_{\rm E} / \omega_{\rm p} \cong 0.49$ \cite{review}, resulting in $\omega_{\rm p} T \approx 4 \pi \approx 12.6$. As a general observation, we can note that the cage correlation functions decay to the 0.1 value -- specified to correspond to the caging time -- on the time scale of 2 -- 5 plasma oscillations for the conditions of Figure~\ref{fig:cagecorr1}. Such a long decay is characteristic for strong-coupled plasmas where the time scale for the diffusion of the particles is slower compared to the plasma oscillations, i.e. the particles are "quasi-localized" on the potential surface \cite{GK2000}. 

We can also observe, by comparing panels (a) and (b) of Figure~\ref{fig:cagecorr1}, that the increasing friction has a more significant effect on the correlation function in the lower-$\beta$ case. At $\beta = 0.5$, the correlation function increases monotonically with increasing $\theta$, however, for $\beta = 1$ a closer observation of the behavior (see the inset in Figure~\ref{fig:cagecorr1}(b)) reveals a non-monotonic behavior. The increasing $\theta$ shifts the crossing of the correlation functions with the $C_{\rm cage}=0.1$ line towards lower times first, and beyond $\theta \approx 0.1$ this trend reverses and remains the same for higher friction values. The panels (c) and (d) of Figure~\ref{fig:cagecorr1} show the same data as in panels (a) and (b), respectively, but with semi-logarithmic scales, to add information about the decay of the cage correlation functions at longer times. The non-monotonicity of the decay of the curves with the friction parameter $\theta$ is also confirmed at longer times (see Figure~\ref{fig:cagecorr1}(d)).  

The effect of a changing strength of magnetic field on the cage correlation functions is presented in Figure~\ref{fig:cagecorr2} for $ \Gamma = 20$ and $\kappa = 2$, for the $\theta=0.1$ (panel (a)) and $\theta=0.5$ values of the friction coefficient. The correlation functions increase monotonically with increasing $\beta$ in both cases, a stronger influence is found at the lower value of friction (panel (a)).   (These data are also shown in semi-logarithmic representation in the panels (c) and (d) of Figure~\ref{fig:cagecorr2}.)

\begin{figure}[h!]
\centering
\includegraphics[width=0.83\textwidth]{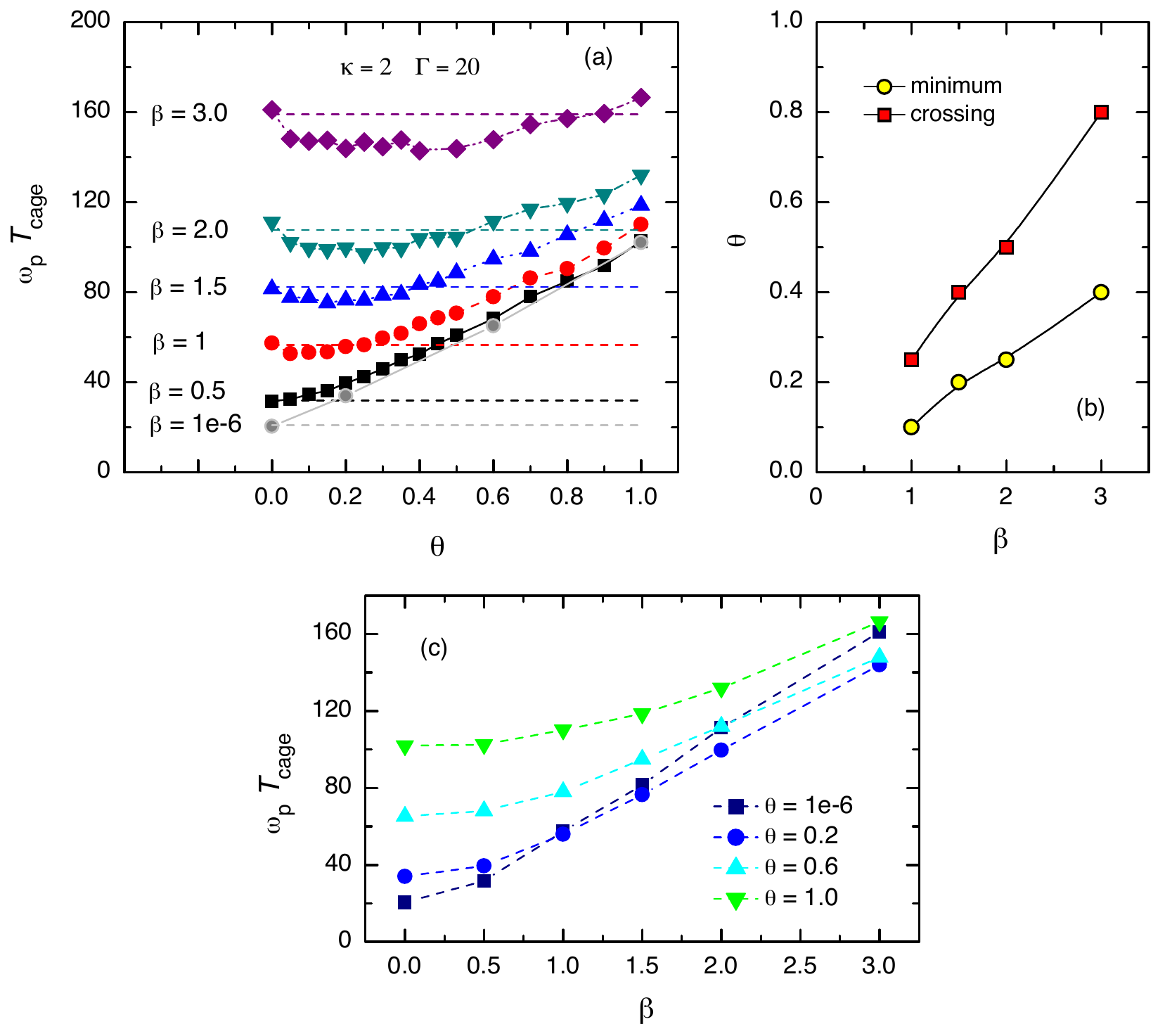}
\caption{\label{fig:cagecorr3}
Dependence of the caging time on the friction parameter $\theta$ at given values of $\beta$, in the highly magnetized domain. Note the non-monotonic dependence of $T_{\rm cage}$ on $\theta$ for the $\beta > 0$ cases. The dashed horizontal lines correspond to $T_{\rm cage}$ at $\theta=0$. (b) Values of the friction coefficient at the minimum of the caging time and at the crossing with the horizontal lines (at $\theta > 0$) in panel (a). (c) Caging time as a function of the normalized magnetic field strength for selected values of the friction coefficient.}
\end{figure}

The non-monotonic dependence of the caging time on the friction coefficient is further analyzed in Figure~\ref{fig:cagecorr3}. At small values of the magnetic field the caging time increases monotonically with increasing friction. At $\beta > 0.5$, however, as already indicated in Figure~\ref{fig:cagecorr1}(b), this dependence is non-monotonic. The effect that the caging time first decreases as a function of $\theta$ gets more pronounced at higher magnetic fields. The value of $\theta$ where the minimum of $T_{\rm cage}$ occurs ($\theta_{\rm min}$), as well as the $\theta$ value ($\theta_{\rm cross}$) where the effect of magnetic field and friction "compensate each other", i.e. when $T_{\rm cage}$ becomes the same again as at $\beta=0$, both increase with increasing magnetic field. These dependences are displayed in Figure~\ref{fig:cagecorr3}(b). Both dependences appear to be nearly linear, the minimum occurs at $\theta_{\rm min} \approx 0.13 \beta$, while the crossing is found at $\theta_{\rm cross} \approx 0.26 \beta$. The dependence of the caging time as a function of $\beta$ is found to be monotonic for any fixed value of the friction coefficient, as it can be seen in Figure~\ref{fig:cagecorr3}(c). 

The interplay of the magnetic field and the friction is non-trivial. Both mechanisms, when acting alone, are known to increase the caging time. The magnetic field results in this by forcing the particles to move on circular trajectories. When the Larmor radius is smaller than the inter-particle separation, diffusive motion across the field lines is significantly hindered and the caging time is enlarged \cite{cagemagnetic}. The effect of friction on the caging time is similar \cite{cagefriction}, but results from a different physical mechanism. As explained earlier, the presence of the gaseous environment is modelled via a damping mechanism by the background (as a continuum) and by a random ("Brownian") force that emulates random kicks by gas particles. The first of these slows down the particle motion, while the second increases the energy of the particle, but randomizes its direction of velocity. The inverse of the frequency $\nu$ (related to the friction coefficient as $\theta = \nu / \omega_{\rm p}$) can be viewed as the timescale for the change of the direction of velocity.

\begin{figure}[h!]
\centering
\includegraphics[width=0.83\textwidth]{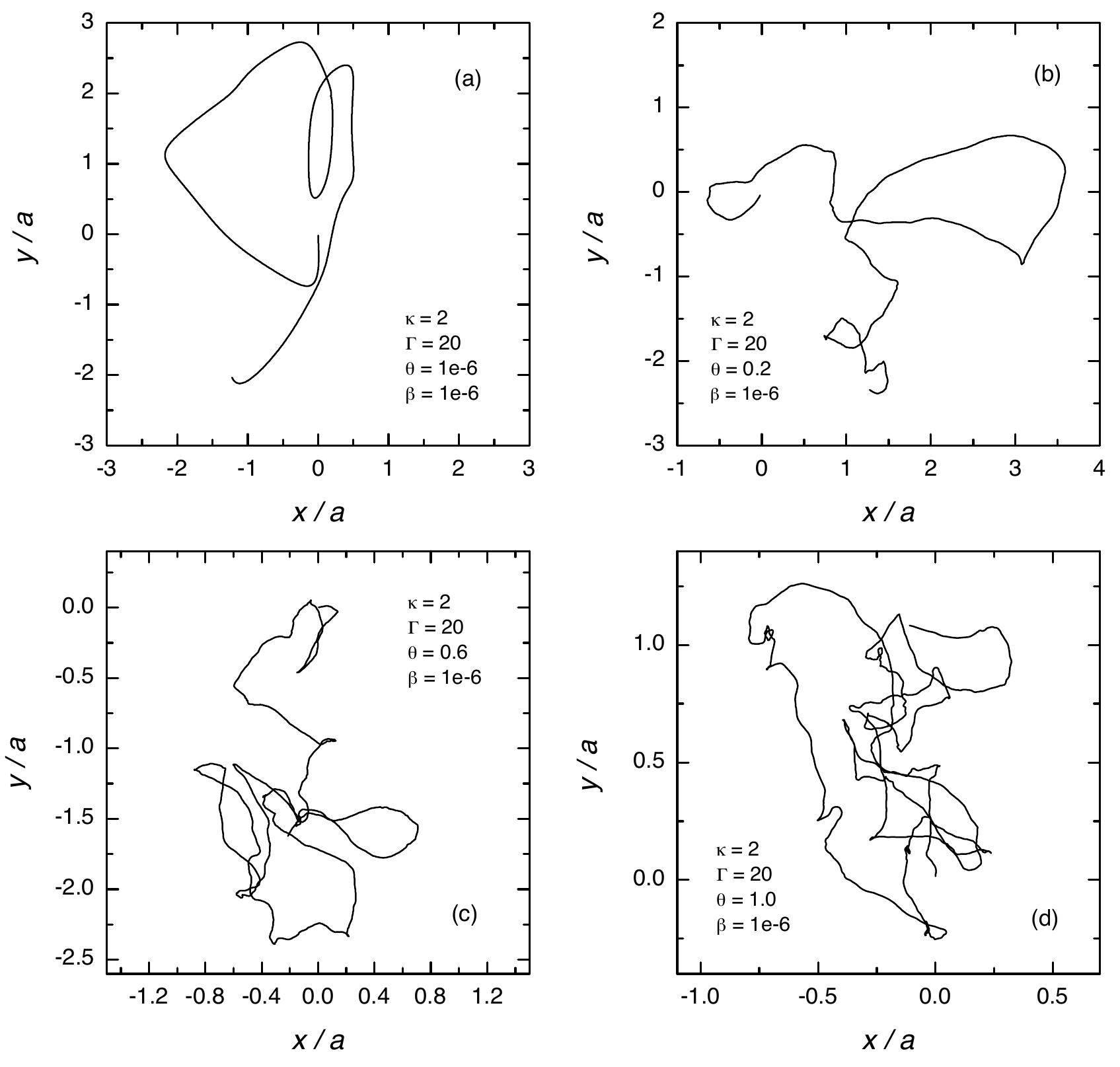}
\caption{\label{fig:cagecorr4}
Trajectories of single particles at different values of the friction parameter, at vanishing magnetic field $\beta=10^{-6}$, $\Gamma=20$ and $\kappa=2$. The length of the trajectory segments is $\omega_{\rm p} T = 200$.}
\end{figure}

\begin{figure}[h!]
\centering
\includegraphics[width=0.83\textwidth]{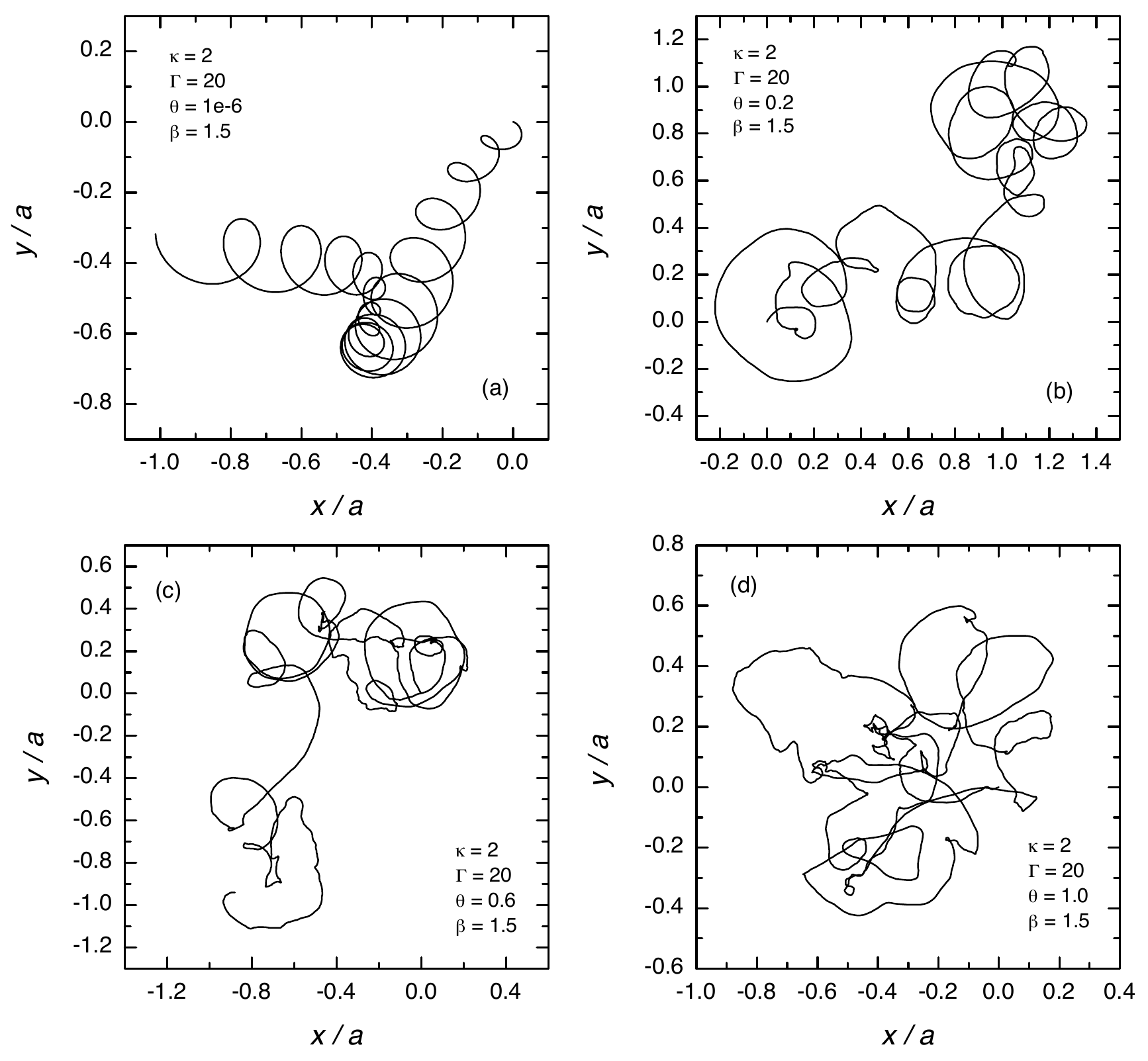}
\caption{\label{fig:cagecorr5}
Trajectories of single particles at different values of the friction parameter, at a magnetic field $\beta=1.5$, $\Gamma=20$ and $\kappa=2$. The length of the trajectory segments is $\omega_{\rm p} T = 200$.}
\end{figure}

Figures~\ref{fig:cagecorr4} and \ref{fig:cagecorr5} show typical particle trajectory segments for various parameter settings. Figure~\ref{fig:cagecorr4} displays these for vanishing magnetic field ($\beta = 10^{-6}$), while in Figure~\ref{fig:cagecorr5} the magnetic field is fixed at $\beta = 1.5$. The four panels of both figures correspond to increasing values of the friction coefficient. The length of recording is $\omega_{\rm p} T = 200$ for all cases. (Moreover, $\Gamma=20$ and $\kappa=2$, as before.)

The trajectory seen in Figure~\ref{fig:cagecorr4}(a) for vanishing friction is smooth, deflections of the velocity are due to inter-particle "collisions". When friction is introduced, the nature of the motion is changed, more frequent changes of direction of motion are apparent already in Figure~\ref{fig:cagecorr4}(b) that corresponds to $\theta=0.2$. With further increasing friction these "breaks" become even more frequent and the particle tends to execute a random walk, which is, however, guided by the neighboring particles -- note the decreasing size of the coordinate domains within which the particle is located during the (constant) time of recording. This is the reason of an increasing caging time with increasing $\theta$.   

Figures~\ref{fig:cagecorr5}(a) displays the trajectory of a single particle for $\beta=1.5$, at vanishing friction. The trajectory consists of smooth "loops" due to the cyclotron motion of the particle combined with diffusion and interaction with the other particles. Such a strong magnetic field enhances significantly the localization of the particle. When the friction is introduced, (see Figure~\ref{fig:cagecorr5}(b), for which $\theta = 0.2$) the circular trajectories (i.e. the loops) are still well visible, however, they are distorted by the random changes of the velocity due to effect of friction. This way the confining effect of the magnetic field is decreased. At further increased friction (distorted) loops are only visible for parts of the trajectory, as seen in Figure~\ref{fig:cagecorr5}(c) for $\theta=0.6$, while they almost completely disappear in Figure~\ref{fig:cagecorr5}(d), which shows the highly damped case of $\theta = 1$. In the latter case the  localization by friction prevails. The case shown in panel (b) corresponds to the minimum of the caging time as a function of $\theta$ (see Figure~\ref{fig:cagecorr3}(a)). For this case $\theta = 0.2$, which corresponds to a "friction" frequency $\nu = 0.2 \omega_p$. The time scale of the motion over one loop, $T$, for this highly magnetized ($\Omega \gg \omega_{\rm E}$) case is $\Omega T = \beta \omega_{\rm p} T \sim 2 \pi$, from which $\nu T \approx \theta (2 \pi / \beta) \approx 0.83$, i.e. the timescale of the change of the direction of motion is approximately the same as the cyclotron period. W suspect that this is the effect that partially destroys of the efficiency of the confinement of the particles by the magnetic field. When the number of collisions per loop doubles (recall that $\theta_{\rm cross} \cong 2 \theta_{\rm min}$), the positive effect of the friction on the localization starts to dominate.

\section{Summary}

In this paper we have investigated the simultaneous effect of friction induced by the gas environment, as well as a homogeneous external magnetic field on the quasi-localization of dust particles in a 2D layer. The system has been described by Langevin Dynamics simulation. We have found that, when acting alone, both an increasing friction coefficient and an increasing strength of the magnetic field enhance the caging of the particles, as quantified by the cage correlation functions. When present simultaneously, however, a non-trivial interplay of the two effects was observed. For a fixed magnetic field ($\beta > 0$) the increasing friction was found first to decrease the caging time and to increase it beyond a certain value of the friction coefficient that depends on the magnetic field strength. A qualitative explanation was given for these observation based on analyzing the peculiarities of the trajectories of individual particles, however a more detailed, quantitative understanding of the effect calls for further studies that include the analysis of the velocity autocorrelation function of the particles. 

\section{Acknowledgment}

This work was supported by the Ministry of Education and Science of the Republic of Kazakhstan within the Program of Grant Funding and the Hungarian National Office for Research, Development and Innovation, via grant NKFIH K-119357.

\end{document}